# NP-Hardness of optimizing the sum of Rational Linear Functions over an Asymptotic-Linear-Program

Deepak Ponvel Chermakani

deepakc@pmail.ntu.edu.sg  deepakc@e.ntu.edu.sg  deepakc@ed-alumni.net  deepakc@myfastmail.com  deepakc@usa.com

*Abstract: -*   We convert, within polynomial-time and sequential processing, an NP-Complete Problem into a real-variable problem of minimizing a sum of Rational Linear Functions constrained by an Asymptotic-Linear-Program. The coefficients and constants in the real-variable problem are 0, 1, -1, K, or -K, where K is the time parameter that tends to positive infinity. The number of variables, constraints, and rational linear functions in the objective, of the real-variable problem is bounded by a polynomial function of the size of the NP-Complete Problem. The NP-Complete Problem has a feasible solution, if-and-only-if, the real-variable problem has a feasible optimal objective equal to zero. We thus show the strong NP-hardness of this real-variable optimization problem.

## 1. Introduction

An Asymptotic-Linear-Program (ALP) is a linear program over real variables, whose coefficients and constants in the objective and constraints, are rational polynomial functions of $K$, the time parameter. It has been proved [1] that as $K$ tends to positive infinity, the ALP demonstrates a steady-state behaviour in its feasibility (or infeasibility) and in its optimal basis of variables (if feasible).

It has been shown [2] that optimizing a single rational polynomial function of real variables, is NP-hard. It has also been shown [3] that optimizing a single rational linear function of binary variables, can be accomplished within polynomial-time.

Consider the problem of optimizing a sum of rational linear functions of real variables, over an ALP. We shall denote this problem as $O_{rational\_linear\_functions\_ALP}$. Denote $P_{rational\_linear\_functions\_ALP}$ as the problem of deciding whether or not the optimal objective value of $O_{rational\_linear\_functions\_ALP}$ is equal to a target integer.

In our paper [4], we showed the NP-Completeness of the problem $P_{linear\_eq\_binary\_1}$ of deciding the feasibility of a set of linear equations over binary variables, with coefficients and constants that are *0, 1*, or *-1*. Consider an instance of problem $P_{linear\_eq\_binary\_1}$ having $M$ linear equations, over a binary variable vector $< b_1, b_2, ... b_N >$, i.e. each variable $b_i$ is allowed to be either *0* or *1*, for all integers $i$ in [*1,N*]:

$a_{1,1} b_1 + a_{1,2} b_2 + ... + a_{1,N} b_N = c_1$
$a_{2,1} b_1 + a_{2,2} b_2 + ... + a_{2,N} b_N = c_2$
…
$a_{M,1} b_1 + a_{M,2} b_2 + ... + a_{M,N} b_N = c_M$

where each of $a_{i,j}$ and $c_i$ is given to be *0, 1*, or *-1*, for all integers $j$ in [*1,N*], and all integers $i$ in [*1,M*]. In the subsequent sections of this paper, we will show how to convert an instance of $P_{linear\_eq\_binary\_1}$ into $P_{rational\_linear\_functions\_ALP}$.

## 2. Modelling Binary Variables using Rational Linear Equations over real variables

**Definition:**   Let $x$ be a real variable such that $0 \leq x \leq 1$. Let $<x_1, x_2, ... x_N>$ be a vector of real variables, such that $0 \leq x_i \leq 1$ for all integers $i$ in [*1,N*]. Let $K$ tend to positive infinity.

**Theorem-1:**  For any positive integer $i$, $( ((x/(K+2i-1)) + ((1-x)/(K+2i))) = 1/(K+2i-x) ) \leftrightarrow (x$ is either *0* or *1*)
**Proof**:   A Boolean statement P↔Q can be proved by showing Q→P and P→Q. For $x=0$, the value of the Left-Hand-Side (LHS) in the equation, is $1/(K+2i)$, which is equal to the value of the Right-Hand-Side (RHS). For $x=1$, the value of the LHS, is $1/(K+2i-1)$, which is equal to the value of the RHS. So Q→P. Next, $((x/(K+2i-1)) + ((1-x)/(K+2i))) = 1/(K+2i-x)$ implies that $( ((x/(K+2i-1)) + ((1-x)/(K+2i))) - 1/(K+2i-x) ) = 0$. Simplifying this expression yields $(x(1-x)) / ((K+2i-1)(K+2i)(K+2i-x)) = 0$. As $K$ tends to positive infinity, the denominator of the LHS of this expression is always positive, so the only way for this equation to be satisfied is that $x(1-x) = 0$, which implies that $x$ is either *0* or *1*. So P→Q.
**Hence Proved**

**Theorem-2:** There is a one-to-one mapping between every $<x_1, x_2, ... x_N>$ and $((x_1/(K+1)) + (x_2/(K+2)) + ... + (x_N/(K+N)))$

**Proof:** Assume that there exists a non-trivial real vector $<\Delta_1, \Delta_2, ... \Delta_N>$ such that $((x_1/(K+1)) + (x_2/(K+2)) + ... + (x_N/(K+N))) = (((x_1 + \Delta_1)/(K+1)) + ((x_2 + \Delta_2)/(K+2)) + ... + ((x_N + \Delta_N)/(K+N)))$. This would imply that $((\Delta_1/(K+1)) + (\Delta_2/(K+2)) + ... + (\Delta_N/(K+N))) = 0$, which would contradict Theorem-1 of the paper [5]. This implies that every real vector $<x_1, x_2, ... x_N>$ corresponds to a unique value of the sum $((x_1/(K+1)) + (x_2/(K+2)) + ... + (x_N/(K+N)))$ and vice-versa.
**Hence Proved**

**Theorem-3:** For each integer $i$ in $[1,N]$, denote $y_i = ((x_i/(K+2i-1)) + ((1-x_i)/(K+2i)))$, and denote $z_i = 1/(K+2i-x_i)$. Then, $(y_1 + y_2 + ... + y_N = z_1 + z_2 + ... + z_N) \leftrightarrow (<x_1, x_2, ... x_N>$ is a binary vector)

**Proof:** A Boolean statement P↔Q can be proved by showing Q→P and P→Q. For $x_i = 0$, the value of $y_i = 1/(K+2i)$, which is equal to the value of $z_i$. For $x_i = 1$, the value of $y_i = 1/(K+2i-1)$, which is equal to the value of $z_i$. So for each element of $<x_1, x_2, ... x_N>$ being either 0 or 1, $(y_1 + y_2 + ... + y_N = z_1 + z_2 + ... + z_N)$. So Q→P. Next, from Theorem-1 of this paper, for any integer $i$ in $[1,N]$, $(y_i = z_i) \rightarrow (x_i$ is either 0 or 1). We now focus on proving that $(y_1 + y_2 + ... + y_N = z_1 + z_2 + ... + z_N) \rightarrow ((y_i = z_i)$, for all integers $i$ in $[1,N]$). Note now from Theorem-2 of this paper (also obvious from Theorem-1 of the paper [5]), that it is not possible for there to exist a non-trivial real vector $<\Delta_1, \Delta_2, ...\Delta_{i-1}, \Delta_{i+1}, ... \Delta_N>$ such that $y_i = \Delta_1 y_1 + \Delta_2 y_2 + ... + \Delta_{i-1} y_{i-1} + \Delta_{i+1} y_{i+1} + ... + \Delta_N y_N$. So coming to the equation $(y_1 + y_2 + ... + y_N = z_1 + z_2 + ... + z_N)$, and considering any $y_i$ in the LHS, the only way to balance it at the RHS, is by $x_i$ taking on a value, such that the denominator of one of the terms of $y_i$ is equal to the denominator of $z_i$. It is not possible for the balancing of $y_i$ to be done by any other $z_j$ ($j \neq i$) because $0 \leq x_i \leq 1$, for all integers $i$ in $[1,N]$. That is either $(K+2i-1) = (K+2i-x_i)$ or $(K+2i) = (K+2i-x_i)$. That is either $x_i = 1$ or $x_i = 0$, and in both these cases, we have $(y_i = z_i)$ as proved in Theorem-1 of this paper. So $(y_1 + y_2 + ... + y_N = z_1 + z_2 + ... + z_N) \rightarrow ((y_i = z_i)$, for all integers $i$ in $[1,N]) \rightarrow (x_i$ is either 0 or 1, for all integers $i$ in $[1,N]$). So P→Q.
**Hence Proved**

**Theorem-4:** The globally minimum value of $(y_1 + y_2 + ... + y_N - z_1 - z_2 - ... - z_N)$ is 0. Also this global minimum is reached when $<x_1, x_2, ... x_N>$ is a binary vector

**Proof:** From Theorem-3 of this paper, it is obvious that $(y_1 + y_2 + ... + y_N - z_1 - z_2 - ... - z_N = 0) \leftrightarrow (<x_1, x_2, ... x_N>$ is a binary vector). We now focus on proving that the minimum value of the expression $(y_1 + y_2 + ... + y_N - z_1 - z_2 - ... - z_N)$ is 0. For any integer $i$ in $[1,N]$, we see that $(y_i - z_i) = ((x_i(1-x_i))/((K+2i-1)(K+2i)(K+2i-x_i)))$. As $K$ tends to positive infinity, and as $x_i$ is constrained between 0 and 1, the denominator of this expression is always positive, and the numerator $(x_i(1-x_i))$ is always non-negative. Hence, the minimum value of $(y_i - z_i)$ is zero, which happens when $x_i$ is either 0 or 1. Thus, the globally minimum value of $(y_1 + y_2 + ... + y_N - z_1 - z_2 - ... - z_N)$ is zero, which happens when $<x_1, x_2, ... x_N>$ is a binary vector.
**Hence Proved**

## 3. Expressing $P_{linear\_eq\_binary\_1}$ as $P_{rational\_linear\_functions\_ALP}$

### 3.1 Obtaining purely Linear constraints, and a sum of Rational Linear Functions for the Objective Function

We use Theorem-4 to express $P_{linear\_eq\_binary\_1}$ as $P_{rational\_linear\_functions\_ALP}$. We aim to minimize $(y_1 + y_2 + ... + y_N - z_1 - z_2 - ... - z_N)$, over the constraints of $P_{linear\_eq\_binary\_1}$, replacing its binary variable vector $<b_1, b_2, ... b_N>$ with the real variable vector $<x_1, x_2, ... x_N>$. If the objective reaches zero, this implies that one of the $2^N$ possible binary vector solutions is allowed for $<x_1, x_2, ... x_N>$. Our intention is to allow the objective of $P_{rational\_linear\_functions\_ALP}$ to have a sum of rational linear functions. We also intend to allow the constraints of $P_{rational\_linear\_functions\_ALP}$ to have purely linear functions (and not rational linear functions). So we make appropriate substitutions for this, and add more linear constraints in the process. For each integer $i$ in $[1,N]$, make the substitution $y_i = (x_i/p_i) + ((1-x_i)/q_i)$, and the substitution $z_i = 1/r_i$, where:

$p_i = (K+2i-1)$
$q_i = (K+2i)$
$r_i = (K+2i-x_i)$

where each of $p_i$, $q_i$, and $r_i$ is a real variable.

Note that the objective is a summation (over all integers $i$ in $[1,N]$) of the term $((x_i(1-x_i))/((K+2i-1)(K+2i)(K+2i-x_i)))$. So we introduce a multiplicative term $K^3$ on the objective. Note that if this multiplicative term is not introduced, any tools that attempt to evaluate the value of the objective will always obtain a value of 0, since the value of $\text{Limit}_{K \rightarrow (\text{positive infinity})}(1/K)$ is considered to be 0. Also note that the value of $(K^3(y_1 + y_2 + ... + y_N - z_1 - z_2 - ... - z_N))$, as $K$ tends to positive infinity, can either be equal to 0, or be equal to a non-zero positive real with a lower bound equal to some function of the coefficients and constants in the linear equations of $P_{linear\_eq\_binary\_1}$ (i.e. it cannot tend to 0 and remain positive).

## 3.2 $O_{rational\_linear\_functions\_ALP}$ and $P_{rational\_linear\_functions\_ALP}$

We write out $O_{rational\_linear\_functions\_ALP}$ with the following Objective and Constraints:

Minimize the Objective:

$K^3$ (
  $(x_1/p_1)$ + $(x_2/p_2)$ + ... + $(x_N/p_N)$
  + $((1-x_1)/q_1)$ + $((1-x_2)/q_2)$ + ... + $((1-x_N)/q_N)$
  - $(1/r_1)$ - $(1/r_2)$ - ... - $(1/r_N)$
)

Subject to Constraints:

$a_{1,1} x_1 + a_{1,2} x_2 + ... + a_{1,N} x_N = c_1$ ;
$a_{2,1} x_1 + a_{2,2} x_2 + ... + a_{2,N} x_N = c_2$ ;
…
$a_{M,1} x_1 + a_{M,2} x_2 + ... + a_{M,N} x_N = c_M$ ;
$0 \leq x_1 \leq 1$;    $p_1 = (K+1)$;    $q_1 = (K+2)$;    $r_1 = (K+2 - x_1)$;
$0 \leq x_2 \leq 1$;    $p_2 = (K+3)$;    $q_2 = (K+4)$;    $r_2 = (K+4 - x_2)$;
...              ...                ...                ...
$0 \leq x_N \leq 1$;   $p_N = (K+2N-1)$; $q_N = (K+2N)$;  $r_N = (K+2N - x_N)$;

$O_{rational\_linear\_functions\_ALP}$ has $3N$ rational linear functions in its objective, $(M+4N)$ linear constraints, and $4N$ real variables. We state $P_{rational\_linear\_functions\_ALP}$ as: (($O_{rational\_linear\_functions\_ALP}$ is feasible) AND (Zero is the minimum objective of $O_{rational\_linear\_functions\_ALP}$)). Finally, ($P_{rational\_linear\_functions\_ALP}$ is TRUE) ↔ (A feasible binary solution exists to $P_{linear\_eq\_binary\_1}$).

## 3.3 Strong NP-hardness of $P_{rational\_linear\_functions\_ALP}$

An ALP whose coefficients and constants are rational functions of $K$, can be expressed with coefficients and constants that are linear functions of $K$. Example, the constraint $(3K^2 + 2K + 5) x < (7/K)$, can be replaced with simultaneous constraints $(y_0 K < 7 ; y_0 = y_1 + y_2 + y_3 ; y_1 = 3K y_{11} ; y_{11} = Kx ; y_2 = 2Kx ; y_3 = 5x)$. Also these constraints may be further expressed with coefficients and constants that are $0, 1, -1, K$, or $-K$. Example, replace $(y_2 = 2Kx)$ with $(y_2 = Kz_1 + Kz_2 ; x = z_1 ; x = z_2)$.

As the maximum magnitude of coefficients in $O_{rational\_linear\_functions\_ALP}$ is $2N$, it can be rewritten (within polynomial time) to have coefficients and constants are $0, 1, -1, K,$ or $-K$. This shows the strong NP-hardness of $P_{rational\_linear\_functions\_ALP}$.

## 4. Conclusion

In this paper, we converted an NP-Complete problem (over binary variables), within polynomial-time, into a decision problem (over real variables) of whether or not the minimum value of a sum of Rational Linear Functions, is zero, constrained by an Asymptotic-Linear-Program. The size (i.e. number of constraints and variables, and rational linear functions in the objective) in the obtained real-variable-problem is bounded by a polynomial function of the size of the given NP-Complete Problem. The real-variable problem can also be efficiently expressed (within polynomial-time) with coefficients and constants that are $0, 1, -1, K,$ or $-K$. We thus, showed that it is strongly NP-hard to optimize the sum of rational linear functions of real variables, constrained by an Asymptotic-Linear-Program.

**About the Author**
I, Deepak Ponvel Chermakani, wrote this paper out of my own interest, during my spare time. In Sep-2010, I completed a fulltime one-year Master Degree in *Operations Research with Computational Optimization*, from University of Edinburgh UK (*www.ed.ac.uk*). In Jul-2003, I completed a fulltime four-year Bachelor Degree in *Electrical and Electronic Engineering*, from Nanyang Technological University Singapore (*www.ntu.edu.sg*). In Jul-1999, I completed fulltime schooling from National Public School in Bangalore in India.